\documentclass[prb,twocolumn]{revtex4}
\usepackage{amsmath}
\usepackage{graphicx}

\begin{document}

\title{Tunneling broadening of vibrational sidebands in molecular transistors}
\author{Karsten Flensberg}
\affiliation{Laboratory of atomic and solid state physics, Cornell University, Ithaca, NY
14853, USA and \O rsted Laboratory, Niels Bohr Institute fAPG,
Universitetsparken 5, 2100 Copenhagen, Denmark.}
\date{\today}

\begin{abstract}
Transport through molecular quantum dots coupled to a single
vibration mode is studied in the case with strong coupling to the
leads. We use an expansion in the correlation between electrons on
the molecule and electrons in the leads and show that the
tunneling broadening is strongly suppressed by the combination of
the Pauli principle and the quantization of the oscillator. As a
consequence the first Frank-Condon step is sharper than the higher
order ones, and its width, when compared to the bare tunneling
strength, is reduced by the overlap between the groundstates of
the displaced and the non-displaced oscillator.

\end{abstract}
\maketitle

\section{Introduction}

In recent years, the field of molecular electronics has renewed
the interest in transport through mesoscopic systems with strong
electron-phonon coupling. A number of experiments have been
reported to demonstrate transport through single
molecules,\cite{reed97,park00,park02,lian02,smit02,zhit02} some of
which show signs of vibrational sidebands. In the pioneering
experiments by Park et al.\cite{park00} it was shown that electron
transport through a single $C_{60}$ molecule was strongly
influenced by a single vibrational mode. The single phonon mode
was associated with the motion of the molecule in the confining
potential created by the van der Waals interactions with the
electrodes. Later similar devices with more complicate molecules
were investigated\cite{park02,zhit02} and they also showed
excitation spectra which possibly could be associated with
emission of vibrational quanta, so-called Frank Condon peaks in
the differential conductance. Some of these
devices\cite{park02,lian02} furthermore exhibited a peak stemming
from the Kondo effect proving that the tunnel coupling to the
leads was rather strong.

Theoretically there has been a large amount of work on the problem
of tunneling through a single level with coupling to phonon modes.
The approaches fall into two categories. The first category is the
kinetic equation approach, which is relevant in the weak tunneling
limit.\cite{rate,flen03} In the kinetic equation approach, it is
essential that the excited vibrational levels are allowed to relax
through coupling to a bath. For a large phonon relaxation rate one
can assume an equilibrium phonon distribution, otherwise a
non-equilibrium distribution function must be determined from the
kinetic equations.\cite{rate} The coupling to the dissipative
environment of the molecule leads to additional broadening of
phonon sidebands.\cite{flen03}

The second category, in which also this paper falls, deals with
the strong tunneling limit. The first approaches were an exact
solution of a simplified situation, where only one electron is
present,\cite{glaz88shek,wing89} i.e., the presence of the Fermi
sea is ignored. The result of this exact solution is that the
phonon sidebands in the tunneling density of states all are
Lorentzians with the same width. The exact solution amounts to
decoupling the electron and oscillator displacement operators.
This decoupling approximation has been used in a number of
papers.\cite{kuochang02,lundin02,alex02} Below we refer to this
approximation as the ``single particle approximation'' (SPA) and
we demonstrate explicitly that this approximation is only valid
for high energies, i.e., for electrons or holes sufficiently far
from the Fermi surface, or in the weak tunneling limit where the
rate equation is valid. For the many-body system many decay
channels are in fact blocked by the Pauli principle. Other
many-body works that include Fermi sea effects have been
done\cite{kral97,koni96,keil02} and the blocking of the line width
has also been alluded to.\cite{kral97,koni96}

In this paper, we focus on the broadening of the Frank-Condon
peaks in the differential conductance due to tunneling broadening,
but in the non-Kondo regime. While this case cannot be solved
exactly, an approximation, which is believed to be valid above the
Kondo temperature and which includes the Pauli principle, but
ignores correlations between lead electrons, is developed. The
main result is that the combination of the Pauli principle and the
splitting of the spectral weight into phonon sidebands severely
limits the phase space for tunnel broadening. Hence the peaks are
much narrower than the bare tunneling rate would suggest. We find
a simple analytic result that describes this, see
Eq.~(\ref{Gappr}) and compare with the SPA formula in
Eq.~(\ref{GRSPAnew}). The approximation is exact in both the
single particle limit and in the weak tunneling limit.
Furthermore, we also support our conclusion by a perturbation
theory in the electron-phonon coupling derived in Appendix
\ref{app:pert}.

\section{Model Hamiltonian and \newline current formula}

\label{sec:model}

We consider a model of a single electron level coupled to two
leads. The single level is coupled to a vibrational mode of the
molecule through the onsite energy. For simplicity we ignore the
spin degree of freedom, which is not relevant unless Kondo-type
effects are important. The model Hamiltonian is
\begin{equation}
H=H_{k}^{{}}+H_{D}^{{}}+H_{DB}^{{}}+H_{B}^{{}}+H_{T}^{{}},\label{Hstart}
\end{equation}
where
\begin{align}
H_{k}^{{}} &
=\sum_{k,\,\alpha=L,R}\xi_{k,\alpha}^{{}}c_{k\alpha}^{\dagger
}c_{k\alpha}^{{}},\quad H_{D}^{{}}=\xi_{0}^{{}}
d_{{}}^{\dagger}d,\\
H_{\mathrm{B}}^{{}} &
=\frac{p_{0}^{2}}{2m_{0}^{{}}}+\frac{1}{2}m_{0}^{{}
}\omega_{0}^{2}x_{0}^{2},\quad H_{DB}^{{}}=\lambda
x_{0}^{{}}d_{{}}^{\dagger }d,\\
H_T^{{}}&=\sum_{k,\,\alpha=L,R}t_{k\alpha}^{{}}
\left(c^\dagger_{k\alpha}d+d^\dagger_{{}}c_{k\alpha}^{{}}\right).
\end{align}
Here $x_{0}^{{}}$ is the oscillator degree of freedom and
$\xi_{0}$ is the bare onsite energy. The lead electron energies
are given by $\xi_{k\alpha}=\varepsilon_{k\alpha}-\mu_{\alpha}$,
where $\mu_{\alpha}^{{}}$ is the chemical potential of the lead
$\alpha$. The tunneling amplitude between lead $\alpha$ and the
molecule is $t_{k\alpha}$. The amplitudes could in principle also
depend on the oscillator position. However, in the experimental
realizations so far this is a negligible dependence and it is
omitted here. Because of this, we can apply the current formula
derived by Wingreen and Meir\cite{meir92} who expressed the
current in terms of the non-equilibrium Green's function of the
molecule. The current through the system is
\begin{equation}
I=\frac{e}{h}\int\frac{d\xi}{2\pi}\frac{\Gamma_{L}^{{}}\Gamma_{R}^{{}}}
{\Gamma_{L}^{{}}+\Gamma_{R}^{{}}}\left[ n_{L}^{{}}(\xi)-n_{R}^{{}}
(\xi)\right]  A(\xi),\label{Igeneral}
\end{equation}
where the left and right distribution functions are
\begin{equation}
n_{\alpha}^{{}}(\varepsilon)=n_{F}^{{}}(\varepsilon-eV_{\alpha}),
\end{equation}
and where $n_{F}^{{}}(\epsilon)=(e^{\beta\epsilon}+1)^{-1}$ is the usual
Fermi-Dirac function. The spectral function, $A$, is given by
\begin{equation}
A(\xi)=-2\operatorname{Im}G^{R}(\xi),
\end{equation}
where the retarded Green's, $G^{R}$, functions are
\begin{equation}
G^{R}(t,t^{\prime})=-i\theta(t-t^{\prime})\langle\{d(t),d_{{}}^{\dagger
}(t^{\prime})\}\rangle.\label{GRdef}
\end{equation}
The widths of the level due to coupling to the leads are
\begin{equation}
\Gamma_{\alpha}^{{}}(\xi)=2\pi\sum_{k}t_{ka}^2\delta(\xi-\xi_{k\alpha}^{{}}).
\end{equation}
The Green's function $G^{R}$ must be calculated in non-equilibrium
and in the presence of the leads.

It is often useful to eliminate the linear coupling terms of the
Hamiltonian Eq.~(\ref{Hstart}) by a unitary transformation similar
to one used in the independent boson model.\cite{mahan} The
transformation is
\begin{equation}
\tilde{H}=SHS^{\dagger},\quad S=e^{-ip_{0}^{{}}\ell d^\dagger d
},\quad \ell=\frac{\lambda}{m_{0}^{{}}\omega_{0}^{2}},
\label{Sdef}
\end{equation}
and with this choice the Hamiltonian transforms into
\begin{equation}
\tilde{H}=H_{k}^{{}}+H_{B}^{{}}+\tilde{H}_{D}^{{}}+\tilde{H}_{T}^{{}
},\label{Huni}
\end{equation}
where
\begin{equation}
\tilde{H}_{D}^{{}}=\varepsilon_{0}^{{}}d_{{}}^{\dagger}d,\quad
\varepsilon_{0}^{{}}=\xi_{0}^{{}}-\frac{1}{2}\lambda\ell,
\end{equation}
and
\begin{equation}
\tilde{H} _{T}=\sum_{k\alpha}t_{k\alpha}^{{}}\left(
c_{k\alpha}^{\dagger}e^{ip_{0}^{{}}\ell}d+d_{{}
}^{\dagger}e^{-ip_{0}^{{}}\ell}c_{k\alpha}^{{}}\right).
\end{equation}
The retarded Green's function becomes after this transformation
\begin{equation}
G^{R}(t-t^{\prime})=-i\theta(t-t^{\prime})\langle\{e^{ip_{0}^{{}}(t)\ell}
d(t),d^{\dagger}(t^{\prime})e^{-ip_0^{{}}(t^{\prime})\ell}\}
\rangle_{S}^{{}},\label{GRS}
\end{equation}
where the average of course should be taken with respect to
$\tilde{H}$, which is indicated by
$\langle\cdots\rangle_{S}^{{}}.$

\section{Calculation of the \newline  spectral function}

In this section, we calculate the spectral function, $A(\xi)$ that
enters into the current formula, Eq.~(\ref{Igeneral}). As
mentioned in introduction, the tunneling broadening of phonon
assisted side bands has been considered before in the case when
the presence of the Fermi sea is ignored(SPA). In this case the
model can be solved exactly, at least in the so-called wide band
limit (WBL). \cite{glaz88shek,wing89} For latter reference we
start by quoting the SPA result
\begin{equation} G^{R,\mathrm{SPA}}(t)=\exp\left(
-t\Gamma/2\right) G_{0} ^{R}(t),\label{GSPA}
\end{equation}
where $G_{0}^R$ is the Green's function in the absence of
tunneling. We see from this expression that the SPA\ implies that
all conductance steps are smeared by the same amount.

\subsection{Dyson equation for $G^R$}

\label{sec:tunnel} In the following, we develop a method to
calculate the broadened Green's function using a truncated
equation of motion technique. Our starting point is the retarded
Green's function in Eq.~(\ref{GRS}). We expand it in terms of
eigenstates of the boson systems
\begin{align}
G^{R}(t)  &  =\sum
G_{nn^{\prime},mm^{\prime}}^{R}(t)f_{nn^{\prime}}
^{{}}f_{mm^{\prime}}^{\ast},\label{GRddef}\\
G_{nn^{\prime},mm^{\prime}}^{R}(t)  & =-i\theta(t)\langle\lbrack
(|n\rangle\!\langle
n^{\prime}|d)(t),d^{\dagger}|m^{\prime}\rangle\!\langle
m|]\rangle_{S}^{{}}, \label{GRdndef}
\end{align}
where we have defined the overlap function between the oscillator
states
\begin{equation}
f_{nn^{\prime}}^{{}}=\langle
n|e^{ip_0^{{}}\ell}|n^{\prime}\rangle.
\end{equation}
An expression for $f_{nn^{\prime}}^{{}}$ is given in
Eq.~(\ref{fnn}).

We now generate a series of equations of motion, starting with one
for $G_{nn^{\prime},mm^{\prime}}^{R}$
\begin{align}
&
(i\partial_{t}+E_{nn^{\prime}}^{{}}-\varepsilon_{0})G_{nn^{\prime
},mm^{\prime}}^{R}(t)=L_{nn^{\prime}}^{R}(t)\nonumber\\
&  +\delta(t)(\langle dd_{{}}^{\dagger}|n\rangle\!\langle
m|\rangle_{S}^{{} }\delta_{n^{\prime}m^{\prime}}^{{}}+\langle
d_{{}}^{\dagger}d|m^{\prime }\rangle\!\langle
n^{\prime}|\rangle_{S}^{{}}\delta_{nm}^{{}}), \label{EOM1}
\end{align}
where $E_{nn^{\prime}}^{{}}=(n-n^{\prime})\omega_{0}^{{}}$, and
$L_{nn^{\prime} }^{R}$ is the contribution stemming from the
tunneling Hamiltonian. The function $L^{R}$ is (see Appendix
\ref{app:EOM} for details of the derivations)
\begin{align}
L_{nn^{\prime}}^{R}(t)  & =\sum_{k\alpha,
l}t_{k\alpha}^{{}}f_{{l}n^{\prime}}^{\ast
}(G_{k\alpha,nl,mm^{\prime}}^{R}(t)-
G_{n_{d}k\alpha,nl,mm^{\prime}}^{R}(t))\nonumber\\
& +\sum_{k\alpha,l}t_{k\alpha}^{{}}f_{nl}^{\ast}
G_{n_{d}k\alpha,{l}n^{\prime},mm^{\prime}}^{R}(t), \label{LRT}
\end{align}
where we encounter two new Green's functions
\begin{align}
G_{k\alpha,nl,mm^{\prime}}^{R}(t)\!\!  &
=\!\!-i\theta(t)\langle\lbrack(|n\rangle\!\langle
n^{\prime}|c_{k\alpha}^{{}})(t),d^{\dagger}|m^{\prime}\rangle\!\langle
m|]\rangle
_{S}^{{}},\label{GRkdef}\\
G_{n_{d}k\alpha,nl,mm^{\prime}}^{R}(t)\! \!&
=\!\!-i\theta(t)\langle\lbrack(|n\rangle\! \langle
n^{\prime}|d^{\dagger}dc_{k\alpha}^{{}})(t),d^{\dagger}|m^{\prime}
\rangle\!\langle m|]\rangle_{S}^{{}}. \label{GRndkdef}
\end{align}
\newline The equation of motion for the first one is
\begin{align}
&
(i\partial_{t}+E_{nl}^{{}}-\xi_{k\alpha}^{{}})G_{k\alpha,nl,mm^{\prime}}^{R}
(t)=L_{k\alpha,nl}^{R}(t)\nonumber\\
&  +\delta(t)(\langle
c_{k\alpha}^{{}}d_{{}}^{\dagger}|n\rangle\!\langle m|\rangle
_{S}^{{}}\delta_{lm^{\prime}}^{{}}+\langle
d_{{}}^{\dagger}c_{k\alpha}^{{} }|m^{\prime}\rangle\!\langle
l|\rangle_{S}^{{}}\delta_{nm}^{{}}), \label{EOM2}
\end{align}
where again $L_{k\alpha,nl}^{R}$ is the contribution from the
tunneling term. Again the tunneling term generates new higher
order Green's functions, but at this point we truncate it using
the following physical principles: we neglect correlations
involving lead electrons, which means that we decouple terms lead
electron operators using the Hartree-Fock approximation.
Furthermore, we set $\langle
c_{k\alpha}^{{}}d^{\dagger}\rangle\approx0$ in the equation of
motion for (\ref{GRkdef}) and (\ref{GRndkdef}). With these
approximations (see Appendix \ref{app:EOM} for more details)
\begin{align}
L_{k\alpha,nl}^{R}(t)  &  \approx t_{k}^{{}}\sum_{j}\langle
n_{k\alpha}^{{}}\rangle
f_{jn}^{{}}G_{jl,mm}^{R}\nonumber\\
&  +t_{k\alpha}^{{}}\sum_{j}(1-\langle
n_{k\alpha}^{{}}\rangle)f_{lj}^{{}}G_{nj,mm}^{R} \label{LTknl}
\end{align}
It is now worth to note that in the case with only one electron,
which is the SPA, the term $\langle n_{k}^{{}}\rangle$ is exactly
zero and Eq.~(\ref{LTknl}) becomes exact for this case. In the
same limit, $G_{n_{d}k\alpha,nl,mm^{\prime}}^{R}=0,$ and the
equations are easily solved. After setting the result back into
Eq.~(\ref{GRddef}) we get
\begin{equation}
G^{R,\mathrm{SPA}}(\omega)=\sum_{nn^{\prime}}|f_{nn^{\prime}}^{{}}
|^{2}\frac{(1-\bar{n})N_{n}^{{}}+\bar{n}N_{n^{\prime}}}
{\omega+E_{nn^{\prime}
}^{{}}-\varepsilon_{0}+i\Gamma/2}, \label{GRSPAnew}
\end{equation}
which is nothing but the SPA result in Eq.~(\ref{GSPA}) written in
the oscillator eigenstate basis. In doing this we have furthermore
evaluated the last term in Eq.~(\ref{EOM1}) as $\langle
dd_{{}}^{\dagger}|n\rangle\!\langle
m|\rangle_{S}^{{}}\delta_{n^{\prime}m^{\prime}}^{{}}+\langle
d_{{}}^{\dagger }d|m^{\prime}\rangle\!\langle
n^{\prime}|\rangle_{S}^{{}}\delta_{nm}^{{}
}=[(1-\bar{n})N_{n}^{{}}+\bar{n}N_{n^{\prime}}]\delta_{n^{\prime}m^{\prime}
}^{{}}\delta_{nm}^{{}},$ where $N_{n}=\langle|n\rangle\!\langle
n|\rangle$ is the occupation of the $n$'th oscillator level and
$\bar{n}=\langle d_{{}} ^{\dagger}d\rangle$ is the average level
occupation.

Here we want to go beyond the SPA and we continue by looking at
the equation of motion for the last Green's function in
Eq.~(\ref{GRndkdef}). We have
\begin{align}
&(i\partial_{t}+E_{nl}^{{}}-\xi_{k\alpha}^{{}})
G_{n_{d}k\alpha,nl,mm^{\prime}}
^{R}(t)=\nonumber\\
& L_{n_{d}k\alpha,nl}^{R}(t)+\delta(t)\langle
c_{k\alpha}^{{}}d_{{}}^{\dagger}|n\rangle\!\langle m|\rangle
\delta_{lm^{\prime}}^{{}}. \label{EOM3}
\end{align}
The function $L_{n_{d}k\alpha,nl}^{R}(t)$ becomes after doing the
same line of approximations as was done for
$L_{k\alpha,nl}^{R}(t)$ (see Appendix \ref{app:EOM})
\begin{equation}
L_{n_{d}k\alpha,nl}^{R}(t)\approx t_{k\alpha}^{{}}\langle
n_{k\alpha}^{{}}\rangle\sum_{j} f_{jn}^{{}}G_{jl,mm^{\prime}}^{R}.
\label{LTndknl}
\end{equation}
The set of equations now close. After integrating out the Green's
function involving the lead electrons, we arrive at linear
equations for the molecule Green's function
\begin{align}
&
(\omega+E_{nn^{\prime}}^{{}}-\varepsilon_{0})G_{nn^{\prime},mm^{\prime}
}^{R}(\omega)=\nonumber\\
&  \phantom{+}\langle dd_{{}}^{\dagger}|n\rangle\!\langle
m|\rangle_{S}^{{} }\delta_{n^{\prime}m^{\prime}}^{{}}+\langle
d_{{}}^{\dagger}d|m^{\prime
}\rangle\!\langle n^{\prime}|\rangle_{S}^{{}}\delta_{nm}^{{}}+\nonumber\\
&  \sum_{lj}\left(
\Lambda_{ln^{\prime}}^{e}f_{nl}^{\ast}f_{jl}^{{}
}G_{jn^{\prime},mm^{\prime}}^{R}(\omega)+\Lambda_{nl}^{h}f_{ln^{\prime}
}^{\ast}f_{lj}^{{}}G_{nj,mm^{\prime}}^{R}(\omega)\right)  ,
\label{Gfinal}
\end{align}
where
\begin{equation}
\Lambda_{nl}^{e,h}(\omega)=\sum_\alpha\int
d\xi\,\frac{\Gamma_\alpha}{2\pi}\left(
\begin{array}
[c]{c}
n_{\alpha}^{{}}(\xi)\\
1-n_{\alpha}^{{}}(\xi)
\end{array}
\right)  \frac{1}{\omega+E_{nl}-\xi+i\eta}. \label{Lambda}
\end{equation}
The functions $\Lambda^{e,h}$ are in the case of energy
independent $\Gamma_\alpha$'s (which we assume henceforth) given
by
\begin{align} \Lambda_{nl}^{e}(\omega)  &
=\sum_\alpha\left[\frac{-i\Gamma_\alpha}{2}n_{\alpha}^{{}
}(\omega+E_{nl}^{{}})+\frac{\Gamma_\alpha^{{}}}{2\pi}
\psi(\omega+E_{nl}^{{}})\right],\nonumber\\
\Lambda_{nl}^{h}(\omega)  &
=\sum_\alpha\left[\frac{-i\Gamma_\alpha^{{}}}{2}n_\alpha^{{}
}(-\omega-E_{nl}^{{}})-\frac{\Gamma_\alpha}{2\pi}
\psi(\omega+E_{nl}^{{}})\right],
\end{align}
where
\begin{equation}
\psi(\varepsilon)=\int d\xi\,n_{F}^{{}}(\xi)\left[
\frac{1}{\varepsilon-\xi }+\frac{1}{\xi}\right]  ,
\end{equation}
where the last term has been added as a regularization of the
integral at large energies. This is allowed because it cancels
when adding the two $\psi$-terms in (\ref{Gfinal}) (which is
easily seen by noting that
$\sum_{l}f_{nl}^{*}f_{jl}^{{}}=\delta_{n,l}$). Note that
$\psi(\varepsilon)$ is a function of $\varepsilon/kT$ only and
that $\psi(\varepsilon)=\psi(-\varepsilon)$. At large values of
$\varepsilon/kT$ the $\psi$-function has the asymptotic form
$\psi(\varepsilon)\approx-\ln(\varepsilon/kT)$, while for small
$\varepsilon$ it goes as
$\psi(\varepsilon)\approx-(\varepsilon/kT)^{2}\times0.213139...$.

The only remaining question is how to evaluate the last term in
Eq.~(\ref{EOM1}). In absence of tunneling or in the single
particle approximation, we can decouple the electron and phonon
degrees of freedom, such that $\langle
dd_{{}}^{\dagger}|n\rangle\!\langle m|\rangle\approx\langle
dd_{{}}^{\dagger}\rangle N_{n}^{{}}\delta_{nm}^{{}},$ where $N_{n}
=\langle|n\rangle\!\langle n|\rangle$ is the occupation of the
$n$'th oscillator level and $\bar{n}=\langle
d_{{}}^{\dagger}d\rangle$ is the average level occupation. To
leading order in $\Gamma$ this is a reasonable approximation.
However, in the more general case we should in principle solve the
terms $\langle dd_{{}}^{\dagger}|n\rangle\!\langle m|\rangle$ and
$\langle d_{{} }^{\dagger}d|m^{\prime}\rangle\!\langle
n^{\prime}|\rangle$ on first right hand of (\ref{Gfinal})
self-consistently. In the general non-equilibrium case, it is,
however, not possible to calculate these expectation values from
the retarded Green's function alone. One must therefore invoke new
approximations for this purpose. Instead of pursuing that, we
concentrate on a case where this is not necessary, namely the
situation of a strongly asymmetric device. This is done in the
following section.

\subsection{Asymmetric transistor, $\Gamma_L\gg\Gamma_R$}

In the strongly asymmetric thermal case, things simplify
considerably, because the dot states are in thermal equilibrium
with one of the leads. For the case $\Gamma_L\gg\Gamma_R$, the
current is
\begin{equation}
I=\frac{e}{h}\int\frac{d\xi}{2\pi}\,\Gamma_{R}^{{}} \left[
n_{L}^{{}}(\xi)-n_{R}^{{}}(\xi)\right]  A_L(\xi),\label{Iasym}
\end{equation}
where $A_L$ is the equilibrium spectral function in equilibrium
with the left lead. Therefore, we can use the standard equilibrium
expression for the self-consistent equations
\begin{align}
\delta_{nm}^{{}}\langle
d_{{}}^{\dagger}d|m^{\prime}\rangle\!\langle n^{\prime }|\rangle &
=\int_{-\infty}^{\infty}\frac{d\omega}{2\pi}A_{nn^{\prime
},mm^{\prime}}(\omega)n_{F}^{{}}(\omega),\\
\delta_{n^{\prime}m^{\prime}}^{{}}\langle dd_{{}}^{\dagger}\rangle
\langle|n\rangle\!\langle m|\rangle &
=\int_{-\infty}^{\infty}\frac{d\omega
}{2\pi}A_{nn^{\prime},mm^{\prime}}(\omega)n_{F}^{{}}(-\omega),
\end{align}
where
$A_{nn^{\prime},mm^{\prime}}(\omega)=i(G_{nn^{\prime},mm^{\prime}
}^{R}(\omega)-G_{d,nn^{\prime},mm^{\prime}}^{A}(\omega))$.
\begin{figure}[ptb]
\setlength{\unitlength}{1cm}
\begin{picture}
(6,9.5)(0,0)
\put(-0.75,0){\includegraphics[width=7cm]{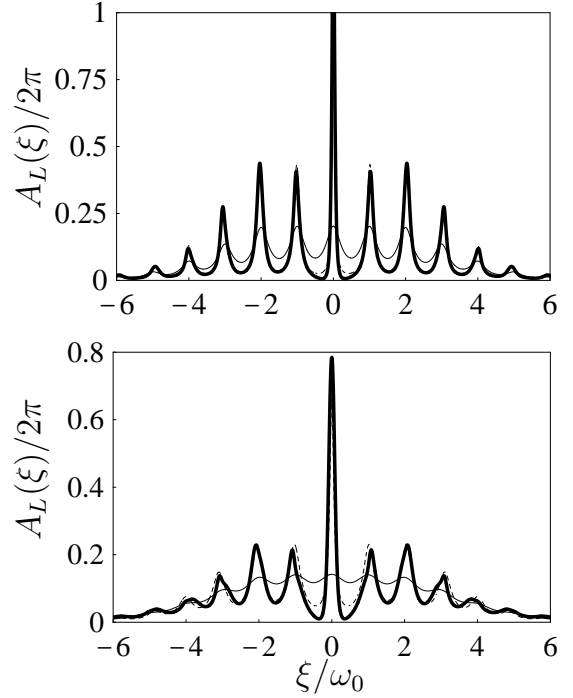}}
\end{picture}
\caption{The broadened tunnel density of states $A_L(\xi)$ for the
resonance condition, $\epsilon_{0}^{{}}=0$, $g=2$ and two
different values of $\Gamma$. Top panel:
$\Gamma/\omega_{0}^{{}}=0.5$ and lower panel: $\Gamma/\omega_{0}
^{{}}=1$. The thick solid lines are the numerical solution of the
full expression, Eq.~(\ref{Gfinal}), while the thin lines are the
single particle approximation. Also shown is the approximation
expression, Eq.~(\ref{Gappr}) with dashed lines. The approximate
expression is in fact bare distinguishable from the full result.
It is clearly seen how the first Frank-Condon step remains sharp
even for large values of tunnel broadening, in accordance with the
estimate in Eq.~(\ref{A00}). In all cases we toke
$kT/\omega_{0}^{{} }=0.01$} \label{fig:tunnel0}
\end{figure}

\begin{figure}[ptb]
\setlength{\unitlength}{1cm}
\begin{picture}
(6,4.5)(0,0)
\put(-.5,-.25){\includegraphics[width=6cm]{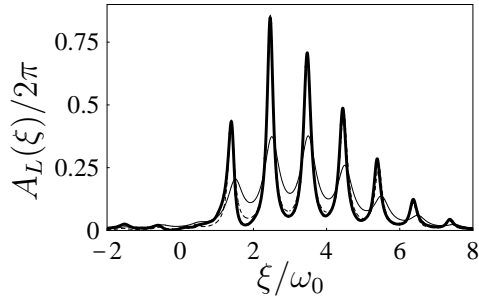}}
\end{picture}
\caption{Same plot as in Fig.~\ref{fig:tunnel0} but with
$\epsilon_{0}^{{} }/\omega_{0}^{{}}=1.5$ and
$\Gamma/\omega_{0}^{{}}=0.5$. Note that when compared to the
$\epsilon_{0}=0$ curve in Fig.~\ref{fig:tunnel0} with the same
$\Gamma$ (top panel) the first step is broader because the
blocking of the level broadening gradually disappears further away
from the Fermi energy.  } \label{fig:tunnel15}
\end{figure}

We have numerically solved these equations and in
Fig.~\ref{fig:tunnel0} we plot the result for the spectral
function, $A_L(\xi)$, for  $\varepsilon_0^{{}}=0$, and different
values of the tunnel broadening. It is clearly seen how the Fermi
surface effects sharpen the lines as compared to the SPA. This
effect is more pronounced for the first peaks, which is also
evident from Fig.~\ref{fig:tunnel15}, where we show the spectral
function for the off-resonance condition.

\subsection{Approximate solution of the Dyson equations}

In order to gain more physical understanding of the narrowing of
the lines seen above and to obtain analytical results, we now
proceed to solve the equations (\ref{Gfinal}) approximately. As is
shown in the numerical plots, the approximate analytic solution,
that is derive below, is in fact close to the full solution.

We look for a solution of Eq.~(\ref{Gfinal}) near one of the
resonances, $\omega\approx\varepsilon_{0}^{{}}+n\omega_{0}$.
Because the Green's function
$G_{d,nn^{\prime},mm^{\prime}}^{R}(\omega)$ peaks at
$\omega=\varepsilon _{0}^{{}}-E_{nn^{\prime}}$, the two terms in
the sum in Eq.~(\ref{Gfinal}) are dominated by,
$E_{jn^{\prime}}^{{}}\approx E_{nn^{\prime}}^{{}}$ and
$E_{nj}^{{}}\approx E_{nn^{\prime}}^{{}}$, respectively. Hence we
can set $j=n$ in the first one and $j=n^{\prime}$ in second one,
which then gives the following diagonal equations
\begin{align}
&
(\omega+E_{nn^{\prime}}^{{}}-\varepsilon_{0}-\Sigma_{nn\prime}^{{}}
(\omega))G_{nn^{\prime},mm^{\prime}}^{R}(\omega)=\nonumber\\
&
\phantom{+}\delta_{nm}^{{}}\delta_{n^{\prime}m^{\prime}}^{{}}\{(1-\bar
{n})N_{n}^{{}}+\bar{n}N_{n^{\prime}}\},
\end{align}
where the self-energy is
\begin{equation}
\Sigma_{nn^{\prime}}^{{}}(\omega)=\sum_{l}\left(
\Lambda_{ln^{\prime}}
^{e}(\omega)|f_{nl}^{{}}|^{2}+\Lambda_{nl}^{h}(\omega)|f_{ln^{\prime}}^{{}
}|^{2}\right)  .\label{selfapp}
\end{equation}
We have, furthermore, approximated
\begin{align}
&  \langle dd_{{}}^{\dagger}|n\rangle\!\langle
m|\rangle_{S}^{{}}\delta _{n^{\prime}m^{\prime}}^{{}}+\langle
d_{{}}^{\dagger}d|m^{\prime}
\rangle\!\langle n^{\prime}|\rangle_{S}^{{}}\delta_{nm}^{{}}\nonumber\\
&
\approx\delta_{nm}^{{}}\delta_{n^{\prime}m^{\prime}}^{{}}\{(1-\bar{n}
)N_{n}^{{}}+\bar{n}N_{n^{\prime}}\},
\end{align}
where $N_{n}=\exp(-n\beta\omega_{0}^{{}})$ and $\bar{n}=\langle
d_{{} }^{\dagger}d\rangle$. This approximation is valid for not
too large $\Gamma$ and we have verified numerically, in the
self-consistent calculation performed above, that it is indeed
reasonable accurate for $\Gamma/\omega_{0}\lesssim 1$. With these
approximations, we obtain a closed expression for the Green's
function $G^{R}$
\begin{equation}
G^{R}(\omega)=\sum_{nn^{\prime}}\frac{(1-\bar{n})N_{n}^{{}}+\bar
{n}N_{n^{\prime}}}{\omega+E_{nn^{\prime}}^{{}}-\varepsilon_{0}
-\Sigma_{nn\prime}^{{}}(\omega)}|f_{nn^{\prime}}^{{}}|^{2}.
\label{Gappr}
\end{equation}

The result in Eq.~(\ref{Gappr}) thus generalizes the SPA\
approximation, Eq.~(\ref{GRSPAnew}), by including the leading
order effect of the Fermi sea. The new self-energy in
Eq.~(\ref{selfapp}) has a simple physical interpretation: the
broadening, which is caused by tunneling out and tunneling in
processes, can only occur if the state in the lead is either empty
or occupied, respectively. For the non-interacting case, where we
in Eq.~(\ref{selfapp}) set $E_{nn'}=0$ and use that
$\sum_{l}|f_{nl}^{{}}|^{2}=1$, the spectral functions reduces
correctly to an Lorentzian with width $\Gamma$.

We can now in detail study the narrowing of the on-resonance line
seen in Fig.~\ref{fig:tunnel0}. At resonance we have
$\bar{n}=\frac{1}{2}$ and we then straightforwardly find
\begin{equation}
A(\omega)\approx\frac{|f_{00}^{{}}|^{2}\Gamma_{00}^{2}}{\omega^{2}
+(\Gamma_{00}^{{}}/2)^{2}},\quad\Gamma_{00}^{{}}=
|f_{00}^{{}}|^{2}\Gamma.
\label{A00}
\end{equation}
This result shows that the width of the resonance at
$\varepsilon_{0}=0$ is considerably narrower than the bare tunnel
broadening would suggest. For large values of $g$
$|f_{00}^{{}}|^{2}$ becomes smaller and the lines narrows in.
However, also the weight of the line goes down with
$|f_{00}^{{}}|^{2}$ as is evident from Eq.~(\ref{A00}).

\subsection{Comparison with perturbation theory in $g$}

It is interesting to compare the results where the tunneling is
treated approximate but the electron-phonon coupling exactly, to
the results of a perturbation theory, where the tunneling is
treated exactly but the electron-phonon coupling to lowest order.
This situation is solved in Appendix \ref{app:pert} and
Fig.~\ref{fig:pert} shows the result of the calculation. Note that
both methods give qualitative the same result, namely that the
zero energy line is narrowed and that spectral function is
suppressed between the two peaks.
\begin{figure}[t] \setlength{\unitlength}{1cm}
\begin{picture}
(6,4)(0,0) \put(0.25,-.25){\includegraphics[width=5cm]{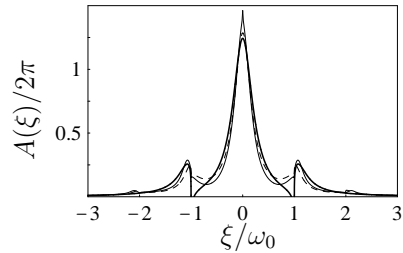}}
\end{picture}
 \caption{The perturbative result for the spectral function
(thick line) compared to the full solution of Eq.~(\ref{Gfinal})
(thin line) and the approximate solution, Eq.~(\ref{Gappr}), for
the symmetric point, $\epsilon_{0}^{{}}=0$. The perturbative
results, which is derived in Appendix \ref{app:pert}, treats the
tunneling term exactly but the electron-phonon coupling only to
linear order. The other parameters are $\Gamma/\omega_{0}
^{{}}=0.5$, $g=0.25$} \label{fig:pert}
\end{figure}

\section{Conclusions}

We have presented results for transport through quantum dots with
strong electron-phonon coupling and with strong tunneling
broadening of the phonon sidebands. This is not a solvable problem
even when the spin degree of freedom is ignored, and we resorted
to an approximation which incorporates the Fermi sea, but ignores
correlation effects. The approximation is exact in both the single
particle and weak tunneling cases. However, physical conclusions
can be drawn from the approximate approach, namely that the tunnel
broadening is much weaker than expected from a model where the
Pauli principle is not incorporated. We have, as mentioned,
neglected the spin degree of freedom which is not expected to be
important at temperatures above the Kondo temperature. It is,
however, an interesting question to ask how the Kondo effect is
influenced by the coupling between the electron occupancy and the
vibrational mode of the molecule.

\acknowledgments The author is thankful for stimulating
discussions with P. Brouwer, A.-P. Jauho, P. McEuen, and J. Sethna
and for financial support from the Danish National Research
Council.

\appendix

\section{Derivation of the equations
of motion} \label{app:EOM}

Here we present the commutators and the methods used to derive the
set of equations of motion in Section \ref{sec:tunnel}. We write
the Hamiltonian in Eq.~(\ref{Huni}) as $H=H_{0}+H_{T}$. In the
oscillator eigenstate representation $H_T$ is
\begin{equation}
H_{T}=\sum_{k\alpha,ll^{\prime}}t_{k\alpha}^{{}}\left(
d_{{}}^{\dagger}c_{k\alpha}^{{} }|l^{\prime}\rangle\!\langle
l|f_{ll^{\prime}}^{\ast}+|l\rangle\!\langle l^{\prime
}|f_{ll^{\prime}}^{{}}c_{k\alpha}^{\dagger}d\right)  .
\end{equation}
\widetext Then Eqs.~(\ref{EOM1}),(\ref{LRT}) follow from
\begin{align}
\lbrack H_{0}^{{}},d|n\rangle\!\langle n^{\prime}|]  &
=(E_{nn^{\prime}
}-\varepsilon_{0}^{{}})d|n\rangle\!\langle n^{\prime}|,\\
\lbrack H_{T}^{{}},d|n\rangle\!\langle n^{\prime}|]  &
=\sum_{k\alpha,ll^{\prime} }t_{k}^{{}}f_{ll^{\prime}}^{\ast}\left(
-d_{{}}^{\dagger}dc_{k\alpha}^{{} }|l^{\prime}\rangle\!\langle
n'|\delta_{nl}^{{}}-dd_{{}}^{\dagger}c_{k\alpha}^{{}
}|n\rangle\!\langle l|\delta_{n^{\prime}l'}^{{}}\right) .
\end{align}
The equation of motion in Eq.~(\ref{EOM2}) follows from the
commutators
\begin{align}
\lbrack H_{0}^{{}},c_{k\alpha}^{{}}|n\rangle\!\langle l|]  &
=(E_{nl}-\xi_{k\alpha}^{{}
})c_{k\alpha}^{{}}|n\rangle\!\langle l|,\label{H0knl}\\
\lbrack H_{T}^{{}},c_{k\alpha}^{{}}|n\rangle\!\langle l|]  &
=\sum_{k^{\prime}\alpha'
jj^{\prime}}t_{k^{\prime}\alpha'}^{{}}\left[
-f_{jj^{\prime}}^{{}}\left(|n\rangle\!\langle j^{\prime}|
\delta_{k\alpha,k^{\prime}\alpha'}^{{}}\delta_{jl}^{{}}+c_{k^{\prime}\alpha'
}^{\dagger}c_{k\alpha}^{{}}(|j\rangle\!\langle
l|\delta_{j^{\prime}n}^{{}}-|n\rangle\!\langle
j^{\prime}|\delta_{jl}^{{}})\right)
d\right.\nonumber\\
&\left. \qquad\qquad +f_{jj^{\prime}}^{\ast}\left(
|j^{\prime}\rangle\!\langle l|\delta_{jn}^{{} }-|n\rangle\!\langle
j|\delta_{j^{\prime}l}^{{}}\right)
c_{k^{\prime}\alpha'}^{{}}c_{k\alpha}^{{}}d_{{} }^{\dagger}\right]
. \label{HTknl}
\end{align}
At this point we truncate the equations of motion approximating
the terms with 3 electron operators by the Hartree-Fock
decomposition for the lead electrons. We have
\begin{align}
c_{k^{\prime}\alpha'}^{\dagger}c_{k\alpha}^{{}}d  & \approx\langle
c_{k^{\prime}\alpha'} ^{\dagger}c_{k\alpha}^{{}}\rangle d-\langle
 c_{k^{\prime}\alpha'}^{\dagger}d\rangle
c_{k\alpha}^{{}},\\
c_{k^{\prime}\alpha'}^{{}}c_{k\alpha}^{{}}d_{{}}^{\dagger}  &
\approx c_{k^{\prime}\alpha'}^{{} }\langle
c_{k\alpha}^{{}}d_{{}}^{\dagger}\rangle -c_{k\alpha}^{{}}\langle
c_{k^{\prime}\alpha' }^{{}}d_{{}}^{\dagger}\rangle.
\end{align}
As mentioned in the main text, we will at this level neglect the
terms $\langle c_{k\alpha}^{{}}d_{{}}^{\dagger}\rangle$ and
therefore only $c_{k^{\prime}\alpha'} ^{\dagger}c_{k\alpha}^{{}}d$
contributes and gives $\langle c_{k^{\prime}\alpha'}^{\dagger
}c_{k\alpha}^{{}}\rangle
d\approx\delta_{k\alpha,k^{\prime}\alpha'}^{{}}\langle
c_{k\alpha}^{\dagger }c_{k\alpha}^{{}}\rangle d$. The last
approximate sign here means up to order $\Gamma$, which is
consistent with neglecting $\langle c_{k\alpha}^{{}}d_{{}
}^{\dagger}\rangle.$ When this is inserted back into
(\ref{HTknl}), we get
\begin{equation} \lbrack
H_{T}^{{}},c_{k\alpha}^{{}}|n\rangle\!\langle
l|]\approx\sum_{jj^{\prime} }t_{k\alpha}^{{}}\left[
-f_{jj^{\prime}}^{{}}\left( |n\rangle\!\langle
j^{\prime}|\delta_{jl}^{{}}+\langle
c_{k\alpha}^{\dagger}c_{k\alpha}^{{}} \rangle(|j\rangle\!\langle
l|\delta_{j^{\prime}n}^{{}}-|n\rangle\!\langle
j^{\prime}|\delta_{jl}^{{}})\right) d\right]  ,
\end{equation}
which then leads to Eq.~(\ref{LTknl}). Finally, the last equation of motion in
Eq.~(\ref{EOM2}) follows from
\begin{align}
\lbrack
H_{0}^{{}},d_{{}}^{\dagger}dc_{k\alpha}^{{}}|n\rangle\!\langle l|]
&
=(E_{nl}-\xi_{k\alpha}^{{}})d_{{}}^{\dagger}dc_{k\alpha}^{{}}|n\rangle\!\langle l|,\\
\lbrack H_{T}^{{}},d_{{}}^{\dagger}dc_{k}^{{}}|n\rangle\!\langle
l|] &
=\sum_{k^{\prime}\alpha',jj^{\prime}}t_{k^{\prime}\alpha'}^{{}}\left[
-f_{jj^{\prime}}^{{} }|j\rangle\!\langle
l|c_{k^{\prime}\alpha'}^{\dagger}c_{k\alpha}^{{}}d\delta
_{j^{\prime}n}^{{}}-f_{jj^{\prime}}^{\ast}|n\rangle\!\langle
j|c_{k\alpha}^{{}}c_{k^{\prime}\alpha'}^{{}}d_{{}}^{\dagger}\delta_{j^{\prime}l}^{{}
}\right]  ,
\end{align}
and after using the same arguments as before, we obtain
\begin{equation}
\lbrack H_{T}^{{}},d_{{}}^{\dagger}dc_{k\alpha}^{{}}
|n\rangle\!\langle l|]\approx -\sum_{j}t_{k}^{{}}f_{jn}^{{}}
|j\rangle\!\langle l|\langle
c_{k\alpha}^{\dagger}c_{k\alpha}^{{}}\rangle d ,
\end{equation}
which leads to Eq.~(\ref{LTndknl}). Furthermore, when the
oscillator states are chosen real the function
$f_{nn^{\prime}}^{{}}$ is
\begin{equation}
f_{nn^{\prime}}^{{}}=\sqrt{\frac{1}{2^{n}2^{n^{\prime}}n!n^{\prime}!}}
e^{-g/2}\left[\mathrm{sign}(n^{\prime}-n)\sqrt{\frac{g}{2}}\,\right]^{|n-n^{\prime}
|}2^{\max(n,n^{\prime})}L_{|n-n^{\prime}|}^{\min(n,n^{\prime})}(g).\label{fnn}
\end{equation}

\section{Perturbation theory in $g$}

\label{app:pert}

In this appendix we derive a perturbative result for the spectral
function to first order in the coupling constant but to all orders
in the tunneling matrix element. We do, however, also assume a
constant bare tunneling density of states and also take the
asymmetric case, $\Gamma_R\ll \Gamma_L=\Gamma$. We want to compute
$G^R$ in Eq.~(\ref{GRdef}) and for the perturbative calculation it
is more convenient to use the Hamiltonian (\ref{Hstart}) instead
of the transformed Hamiltonian. We will start from the equations
of motion for the Green's function $G^R_d$. We use here the
notation $G_{y}^{R}=-i\theta(t)\langle\lbrack
y(t),d^{\dagger}]\rangle.$ We obtain
\begin{equation}
\left(  \omega-\xi_{0}+i\frac{\Gamma}{2}\right)
G_{d}^{R}(\omega)=1+\lambda G_{xd}^{R}(\omega). \label{Gd}
\end{equation}
The equation of motion for the function $G_{xd}^R$ is
\begin{equation}
\left(
\begin{array}
[c]{cc}
\omega-\xi_{0} & -i/m\\
im\omega_{0}^{2} & \omega-\xi_{0}
\end{array}
\right)  \left(
\begin{array}
[c]{c}
G_{xd}^{R}\\
G_{pd}^{R}
\end{array}
\right)  =\left(
\begin{array}
[c]{c}
\langle x\rangle\\
0
\end{array}
\right)  +\sum_{k}t_{k}\left(
\begin{array}
[c]{c}
G_{xk}^{R}\\
G_{pk}^{R}
\end{array}
\right)  +\frac{\lambda}{2}\left(
\begin{array}
[c]{c}
\ell_{0}^{2}\\
-i
\end{array}
\right)  G_{d,\lambda=0}^{R}(\omega), \label{Gxdpd}
\end{equation}
and
\begin{equation}
\left(
\begin{array}
[c]{cc}
\omega-\xi_{k}+i\eta & -i/m\\
im\omega_{0}^{2} & \omega-\xi_{k}+i\eta
\end{array}
\right)  \left(
\begin{array}
[c]{c}
G_{xk}^{R}\\
G_{pk}^{R}
\end{array}
\right)  =t_{k}\left(
\begin{array}
[c]{c}
G_{xd}^{R}\\
G_{pd}^{R}
\end{array}
\right)  -i\lambda\left(
\begin{array}
[c]{c}
0\\
1
\end{array}
\right)  G_{nk,\lambda=0}^{R}(\omega), \label{Gxkpk}
\end{equation}
where we used that since $G^R_{xd}$ in (\ref{Gd}) is multiplied by
$\lambda$, the last term of (\ref{Gxkpk}) should be calculate
without electron phonon coupling.  Therefore, for temperatures
$kT\ll\omega_0$ the following replacements were made in that term
$ \lambda x^{2}d=\lambda\langle
x^{2}\rangle_{0}^{{}}d=\frac{1}{2}\ell _{0}^{2}d $, and $ \lambda
pxd=\lambda\langle px\rangle_{0}^{{}}d=\frac {-i}{2}d$. The
average value $\langle x\rangle$ appearing in Eq.~(\ref{Gxdpd})
follows from $\langle\ddot{x}\rangle=-\omega_{0}^{2}\langle
x\rangle-\langle n_{d} ^{{}}\rangle\lambda/m=0$. Expressing the
functions $G^R_{xk}$ and $G^R_{pk}$ in Eq.~(\ref{Gxkpk}) in terms
of $G^R_{xd}$ and $G^R_{pd}$ and inserting this back into
Eqs.~(\ref{Gxdpd}) and (\ref{Gd}),  we obtain
\begin{equation}
G_{d}^{R}(\omega)=\frac{1-g}{N_{0}}-\frac{g\omega_{0}}{N_{0}^{2}}
+\frac{gn}{N_{+1}}+\frac{g\left(  1-n\right)
}{N_{-1}}+\frac{\lambda^{2}}{m}\frac{1}{N_{+1}
N_{-1}N_{0}}\sum_{k}\frac{t_{k}^{{}
}\left(  N_{0}+\omega-\xi_{k}\right)  }{\left(
\omega-\xi_{k}+i\eta\right)
^{2}-\omega_{0}^{2}}G_{nk,\lambda=0}^{R}(\omega), \label{Gdapp}
\end{equation}
where $N_{p}=\omega-\xi_{0}+p\omega_{0}+i\frac{\Gamma}{2}$. The
first four terms in (\ref{Gdapp}) are nothing but the SPA\ result
in Eq.~(\ref{GRSPAnew}) expanded to lowest order in $g$.
\emph{Therefore, it is evident that the last term of
Eq.~(\ref{Gdapp}) constitutes the correction to the SPA}. The
Green's function $G_{nk,\lambda=0}^{R}$ can be evaluated using
Wick's theorem
\begin{equation}
G_{nk,\lambda=0}^{R}(\omega)=\langle
n_{d}\rangle_{0}^{{}}G_{k,\lambda=0} ^{R}(\omega)-\langle
d^{\dagger}c_{k}\rangle_{0}^{{}}G_{d,\lambda=0} ^{R}(\omega),
\label{GRnkl0}
\end{equation}
where
\begin{equation}
G_{k,\lambda=0}^{R}(\omega)=t_{k}\frac{1}{\omega-\xi_{k}+i\eta}G_{d}
^{R}(\omega), \quad \langle
d^{\dagger}c_{k}\rangle_{0}^{{}}=i\int\frac{d\omega^{\prime}}{\pi
}n_{F}^{{}}(\omega^{\prime})\left(
G_{k,\lambda=0}^{R}(\omega^{\prime
})-G_{k,\lambda=0}^{A}(\omega^{\prime})\right).\label{Gnuller}
\end{equation}
The integrant in Eq.~(\ref{Gdapp}) is analytic in the upper half
plane of the complex $\xi_{k}$ plane, and when performing the
integral by a contour in the lower half-plane only one term in
Eq.~(\ref{Gnuller}) contributes. The last term in
Eq.~(\ref{Gdapp}), which was the correction to the SPA result,
then reads
\begin{equation}
\delta
G_{d}^{R}(\omega)=\frac{-g\omega_{0}^{2}\Gamma}{N_{+1}N_{-1}N_{0}^{2}
}\gamma(\omega) \label{dGf}
\end{equation}
where
\begin{equation}
\gamma(\omega)=\omega_{0}\int_{-\infty}^{\infty}\frac{d\omega^{\prime}}{\pi
}n_{F}^{{}}(\omega^{\prime})\frac{1}{\omega^{\prime}-\xi_{0}^{{}}-i\Gamma
/2}\frac{2\omega-\xi_{0}-\omega^{\prime}}{\left(  \omega^{\prime}-\omega
-i\eta\right)  ^{2}-\omega_{0}^{2}}.
\end{equation}
The SPA\ is restored when the energy is either far above or far
below the Fermi surface, because the model is then similar to a
single electron or a single hole. This is clearly seen in the
limit $|\xi_{0}|,\omega_{0}\gg\max (\Gamma,kT)$, where we can
replace $n_{F}$ by either one or zero. For the first cases it is
clear that $\delta G$ is 0. When $n_{F}=1$, we have also get zero
because the integrant is analytic in one half plane.

Finally, at $T=0,$ we have
\begin{align*} \gamma(\omega)  &
=\frac{\omega_{0}}{2\pi}\frac{2\omega-2\xi_{0}^{{}}
-i\Gamma/2}{(\omega-\xi_{0}^{{}}-i\Gamma/2)^{2}-\omega_{0}^{2}}\left(
2i\pi n_{0}^{{}}+\ln\left[
(\xi_{0}^{{}}/\omega_{0}^{{}})^{2}+(\Gamma/2\omega
_{0}^{{}})^{2}\right]  \right) \\
&  +\frac{\omega-\xi_{0}-\omega_{0}^{{}}}{2\pi(\omega-\xi_{0}^{{}}+\omega
_{0}^{{}}-i\Gamma/2)}\left(  i\pi\theta(-\omega_{0}-\omega)+\ln\left|
(\omega_{0}^{{}}+\omega)/\omega_{0}^{{}}\right|  \right) \\
&  -\frac{\omega-\xi_{0}+\omega_{0}^{{}}}{2\pi(\omega-\xi_{0}^{{}}-\omega
_{0}^{{}}-i\Gamma/2)}\left(  i\pi\theta(\omega_{0}-\omega)+\ln\left|
(\omega_{0}^{{}}-\omega)/\omega_{0}^{{}}\right|  \right)  ,
\end{align*}
where
\begin{equation}
n_{0}^{{}}=\frac{1}{\pi}\tan^{-1}\left(  \frac{-2\xi_{0}^{{}}}{\Gamma}\right)
+\frac{1}{2}.
\end{equation}
The zero temperature expression for the SPA\ correction is seen to
diverge at $|\omega|=\omega_{0}$. This divergence is, however,
cut-off at finite temperatures. If Fig.~\ref{fig:pert} we show the
result of the perturbation theory compared to the approximate
spectral function derived in the main text. The plot is at the
symmetric point $\varepsilon
_{0}^{{}}=\xi_{0}-g\omega_{0}^{{}}=0$. One should note that when
expanding $\varepsilon_{0}^{{}}$ to linear order it cancels off
the term $-g\omega_{0}^{{}}/N_{0}^{2}$ in Eq.~(\ref{Gdapp}).


\end{document}